\documentclass[12pt]{iopart}

\def\undersymbol#1#2{\mathop{\vtop{\ialign{##\crcr
     $\hfil\displaystyle{#2}\hfil$\crcr\noalign
     {\kern1pt\nointerlineskip}\hbox{$\hfil#1\hfil$}\crcr
     \noalign{\kern1pt}}}}}

\usepackage{graphicx}
\begin{document}

\title{POLARIZATION IN BINARY MICROLENSING EVENTS}

\author{G. Ingrosso, F. De Paolis, A. A. Nucita, F. Strafella}
\address{Dipartimento di Matematica e Fisica "E. De Giorgi", Universit\`a del Salento, 
Via per Arnesano, I-73100, Lecce, Italy and \\
INFN, Sezione di Lecce, Via per Arnesano, I-73100, Lecce, Italy}
\ead{ingrosso@le.infn.it}

\author{S. Calchi Novati} 
\address{Dipartimento di Fisica ``E. R. Caianiello'', 
Universit\`a di Salerno, I-84084 Fisciano (SA), Italy and \\
Istituto Internazionale per gli Alti Studi Scientifici (IIASS), 
Vietri Sul Mare (SA), Italy} 

\author{Ph. Jetzer}
\address{Institute f\"ur Theoretische Physik, Universit\"at Z\"urich, Winterthurerstrasse
190, 8057 Z\"urich, Switzerland}

\author{G. Liuzzi} 
\address{Scuola di Ingegneria, Universit\`a degli Studi della Basilicata, 
via dell'Ateneo Lucano 10, 85100, Potenza, Italy} 

\author{A. Zakharov} 
\address{Institute of Theoretical and Experimental Physics, B. Cheremushkinskaya 25, 
117259 Moscow,  Russia and \\
Bogoliubov Laboratory of Theoretical Physics, Joint Institute for Nuclear Research, 
141980 Dubna, Russia and \\
North Carolina Central University, 1801 Fayetteville Street, NC 27707 Durham, USA}

\begin{abstract}

The light received by source stars in microlensing events may be significantly polarized if both an efficient 
photon scattering mechanism is active in the source stellar atmosphere and a differential magnification is therein 
induced by the lensing system. The best candidate events for observing polarization are highly magnified events  
with source stars belonging to the class of cool, giant stars {in which the stellar light is polarized by 
photon scattering on dust grains contained in their envelopes. The presence in the stellar atmosphere of an internal 
cavity devoid of dust produces polarization profiles with a two peaks structure. Hence, the time interval
between them gives an important observable quantity directly related to the size of the internal cavity 
and to the model parameters of the lens system.}
We show that {during a microlensing event} the expected polarization variability can solve an ambiguity, 
that arises in some cases, related to the binary or planetary lensing interpretation of the perturbations 
observed near the maximum of the event light-curve. 
We consider a specific event case for which the parameter values corresponding to the two solutions are given.
Then, assuming a polarization model for the source star, we compute the two expected polarization profiles.
The position of the two peaks appearing in the polarization curves and the characteristic time interval between 
them allow us to distinguish between the binary and planetary lens solutions.

\end{abstract}

\section{Introduction}

Gravitational microlensing technique initially developed to search 
for MACHOs (Massive Astrophysical Compact Halo Objects) in the Galactic halo  
by long observational {campaigns} towards several directions in the sky
\cite{Macho93} \cite{Eros93} \cite{Ogle94} \cite{Sebastiano10},
has become nowadays a powerful tool to investigate several astrophysical phenomena.
Microlensing observations have been used: \\
$-$ to map the amount and distribution of luminous matter in the Galaxy, Magellanic Clouds and M31
galaxy \cite{GYUK1999,EVANS2002,MONIEZ2010}; \\
$-$ to carry out detailed studies of different classes of variable stars which actually do change their brightness due to changes in size and temperature \cite{OGLE2011,OGLE2013}; \\
$-$ to test stellar atmosphere models via the detection of limb-darkening effects
\cite{GG1999,ABE2003};\\
$-$ to discover and fully characterize exoplanetary systems, via the detection
of anomalies in the microlensing light-curves expected for single-lens events.
{ Indeed, up to now, 22 planetary systems in the Galaxy 
have been discovered with this technique (see http://exoplanet.eu).}
Moreover, the anomaly in pixel lensing found in \cite{An04} can be explained with an exoplanetary system
in the M31 galaxy \cite{Ingrosso09,Ingrosso11}.

In the present paper we consider the possibility that 
during a microlensing event, depending on the nature of the
source star and the parameters of the microlensing event,
a characteristic polarization signal of the source star light might arise.
It has already been shown that polarization measurements offer an unique opportunity
to probe stellar atmospheres of very distant stars and also to measure 
the lens Einstein radius $R_{\rm E}$, if the physical radius $R_S$ of the source is known.
Moreover, since the polarization curve is sensitive to the presence of lens planetary companions, 
polarization measurement may help to { retrieve}  the parameters of the binary-lens system.

Polarization of the stellar light is caused by photon scattering in the stellar atmospheres
of several classes of stars. In particular: \\
$-$ in the case of hot stars (O, A, B type), light is polarized
by Thomson scattering on free electrons. 
This phenomenon has been completely studied by Chandrasekhar 
\cite{Chandra60}, showing that
the linear
polarization increases from the center to the star limb, where about 12 per cent of the light
is polarized. However, hot stars are rather rare and, indeed, no source star
of this type has been observed as source in microlensing events. \\
$-$ in the case of main sequence stars of late type (G, K, M), light is
polarized by the coherent Rayleigh scattering on neutral hydrogen and molecules \cite{Stenflo}.
These stars constitute the larger fraction of the source stars in microlensing events,
but the polarization degree is lower (about 3 order of magnitude) with respect to hot star case. \\ 
$-$ in the case of cool giant stars, the stellar light is polarized by photon scattering on
dust grains contained in their envelopes powered by large stellar winds 
\cite{Simmons95,Simmons02,Ignace06}.

Cool giant stars constitute a significant fraction of the lensed sources in microlensing events
towards the Galactic bulge, the LMC  and the M31 galaxy. Moreover, 
the polarization signal is expected to be relevant, particularly for red giants having 
large dust optical depth. These source stars are the more valuable 
candidates for observing a polarization signal during a microlensing event
{ with source stars in the Galactic bulge.}
In this paper we concentrate particularly on this kind of sources.

A variable polarization across the stellar disk  is currently 
observed only for the Sun \cite{Stenflo} and, as expected,
the polarization degree increases from the  center
to the star limb. In the case of distant stars, the stellar disk is not 
resolved and only the overall polarization is relevant. This
is usually zero, since the flux from each stellar disk element is the same.
A net polarization of the stellar light is produced if some suitable asymmetry 
is present in the stellar disk
\footnote {Polarization is also produced in the propagation of the stellar light through the interstellar medium.
This contribution to the total effect is to be subtracted in real observations.}
due, e.g., to hot spots, tidal distorsions, eclipses, 
fast rotation or magnetic fields.

In the microlensing context, an overall polarization of the stellar light 
is always present since different parts of the source star disk are differently magnified
by the lens system. Moreover, due to the relative motion between source and lens, 
the gravitational lens scans the disk of the source star giving rise also to a time dependent 
polarization signal.
The polarization signal will be relevant, and possibly observable, in events with 
high magnification (both single lens and binary), which also show large finite source effects,
namely for events in which the source star radius is of the order or greater than the lens impact parameter.

In a recent work \cite{Ingrosso12} we considered a specific set of highly magnified, 
single-lens events and a subset of exoplanetary events observed towards the Galactic bulge.
 As an illustration, we also considered the expected polarization signal
for the PA-99-N2 exoplanetary event towards M31.
We calculated the polarization profiles as
a function of time taking into account the nature of the source stars.
Given a $I$ band typical magnitude at maximum
magnification of about 12 and a duration
of the polarization signal up to 1 day, we showed 
that the currently available technology,
in particular the polarimeter in FORS2 on the VLT, 
may  potentially allow the detection of such signals.

Besides the interest related to stellar astrophysics, the 
analysis  of a polarization profile (which is related to the underlying 
magnification light-curve) may in principle provide
independent constraints on the lensing parameters in binary events.
The aim of the present paper is to show that, given sufficient observational precision,
polarization measurements are able to solve a specific type of ambiguity, namely the planet or binary 
interpretation of anomalies present in microlensing light curves.
{The general method is similar to that of \cite{Loeb-Sasselov} in which the presence 
in giant stars of resonant lines with intensity increasing from the center to the star limb
(and a variable magnification across the stellar disk) 
leads to narrow band (centered on the resonance line) stellar fluxes with a two peaks structure. 
Similarly, we obtain polarization profiles 
with a double peak structure and the observable time interval 
between them becomes 
an important tool to investigate both the source and lens parameters.}

High magnification microlensing events provide an important channel to detect planets, via
the detection of perturbations near the peak of the events. 
It is known that these perturbations can be produced by a planet or a binary companion to the primary lens and 
that both types of solutions can be generally distinguished, due to different 
magnification patterns around caustics. However, there are cases (that are expected to be common),
in which the degeneracy between the planet and binary solution cannot be resolved 
by the analysis of the light curves. We consider in particular
the OGLE-2011-BLG-0905/MOA-2011-BLG-336 
event case \cite{Choi12} and we show that the expected polarization curves are different for the
planet and binary case, potentially allowing to solve the ambiguity.
Of course, since accurate polarization measurements cannot be obtained 
with a survey telescope, alert systems are necessary allowing large 
telescopes to take polarimetry measurements during a microlensing event.

\section{Generalities}

Following the approach outlined in \cite{Chandra60}
we define the intensities $I_l(\mu)$ and $I_r(\mu)$ emitted by the scattering 
atmosphere in the direction making an angle $\chi$ with the 
normal to the star surface and polarized as follows: 
$I_l(\mu)$ is the intensity in the plane containing 
the line of sight and the normal, 
$I_r(\mu)$ is the intensity in the direction perpendicular to this plane.

{
We choose a coordinate system in the lens plane with the 
origin at the center of mass of the binary system. 
The $Oz$ axis is directed towards the observer, 
the $Ox$ axis is oriented parallel to the binary component separation.
The location of a point $(x,y)$ on the source star surface is determined by the 
angular distance $\rho$ from the projected position of the source star center
$(x_0,y_0)$ and by the angle $\varphi$ with the $Ox$ axis
($x=x_0 + \rho \cos \varphi$ and $y=y_0 + \rho \sin \varphi$).
In the above coordinate system $\mu = \sqrt{1-\rho^2/\rho^2_S}$, 
where $\rho_S$ is the angular source radius.
Here and in the following all angular distances 
are given in units of the Einstein angular radius $\theta_E$ 
of the total lens mass.}

To calculate the polarization of a star 
we integrate the unnormalized Stokes parameters and the flux over the star disk
\cite{Simmons95,Agol96}
\begin{eqnarray}
F = F_0 \int_{0}^{2\pi}   \int_0^{\rho_S} A(x,y) ~ 
I_+(\mu) ~\rho d\rho~ d \varphi~,
\label{flux} 
\end{eqnarray}
\begin{eqnarray}
F_Q = F_0 \int_{0}^{2\pi} \int_0^{\rho_S} A(x,y) ~ 
I_-(\mu)~\cos 2 \varphi ~\rho d\rho~ d \varphi~,
\label{fq} 
\end{eqnarray}
\begin{eqnarray}
F_U = F_0 \int_{0}^{2\pi} \int_0^{\rho_S} A(x,y) ~ 
I_-(\mu) ~ \sin 2 \varphi ~\rho d\rho~ d \varphi~,
\label{fu} 
\end{eqnarray}
where $F_0$ is the unmagnified star flux, 
$A(x,y)$ is the point source 
magnification due to the lens system and 
$I_+(\mu) = I_r(\mu)+I_l(\mu)$ and $I_-(\mu)=I_r(\mu)-I_l(\mu)$
are intensities related to the considered polarization model.

As usual \cite{Chandra60}, the polarization degree is  
$P = (F_Q^2+F_U^2)^{1/2}/F$ and 
the polarization angle
$\theta_P = (1/2) \tan^{-1} (F_U/F_Q)$.

Since we are dealing with binary events for which the source trajectory may intersect either fold 
{caustics} or cusps (where the lensing magnification of a point source becomes 
infinite), instead of directly solving the lens system equations \cite{WittMao},
we evaluate the magnification $A(x,y)$ at any point in the source plane 
by using the Inverse Ray-Shooting method \cite{Kayser,Wambsganss}. 
As it is well known, the magnification depends on the mass ratio $q$ between the binary 
components and on their projected separation $d$ (in units of the Einstein radius $R_E$).

{ Further parameters entering in the above equations are 
the coordinates $(x_0,y_0)$ 
of the source star center.} These are given, at any time $t$, in terms of the other lens system parameters, that are: 
the maximum amplification time $t_0$, the impact parameter $u_0$ (which is the minimum distance between source star center 
and the center of mass of the lens system), the Einstein time $t_{\rm E}$ and the angle $\alpha$ of the source 
trajectory with respect to the $Ox$ axis connecting the binary components.  

\section{Polarization for cool giant stars}

Polarization during microlensing of source stars with extended envelopes has been studied  
for  single-lens events \cite{Simmons02} and for binary lensing \cite{Ignace06}. 
As emphasized in these works, the model is well suited to describe
polarization in evolved, cool stars that exhibit stellar winds  
significantly stronger than that of the Sun.

The scattering opacity responsible for producing the polarization is
the photon scattering on dust grains. However, 
since the presence of dust is only possible at radial distances
at which the gas temperature is below the 
dust sublimation temperature $T_h \simeq 1300$ K (depending on the grain 
composition), in our model a circumstellar cavity is considered between the 
photosphere radius and the condensation radius $R_h$, where the 
temperature drops to $T_h$. 

The dust number density distribution is 
parametrized by a simple power law 
\begin{eqnarray}
n_{dust}(r) = n_h ~ (R_h/r)^{\beta}~~~~~~{\rm for}~~~ r>R_h~,
\label{density_law}
\end{eqnarray}
where $r = (\rho^2+z^2)^{1/2}$ is the radial distance
from the star center, $n_h$ is the dust number density at
the radius $R_h$ of the central cavity and $\beta$ is a free parameter
depending on the velocity structure of the wind: 
$\beta=2$ holds for constant velocity winds while larger values correspond to
accelerated winds.

\begin{figure}[htbp]
\vspace{5.5cm} \includegraphics{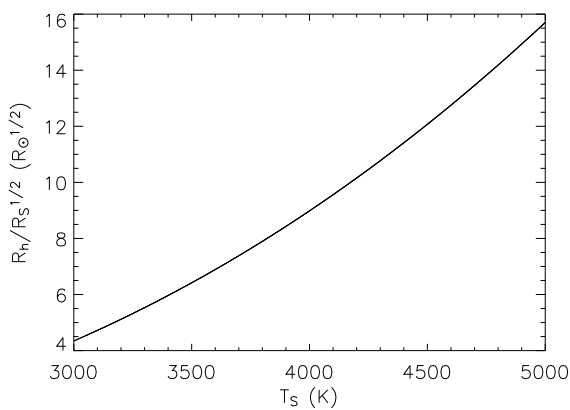}
\caption{The { ratio $R_h/\sqrt{R_S}$ is} shown as a function of the stellar surface temperature $T_S$. 
The typical dust sublimation temperature $T_h \simeq 1300$ K has been assumed.}
\label{fig1}
\end{figure}

We estimate $R_h$ according to simple energy balance 
criteria. We consider the balance between the energy 
absorbed and emitted by a typical dust grain as a function of the radial distance from the star
\begin{equation}
\int_0^\infty F^S_\lambda(r) \pi a^2 Q_\lambda d\lambda= \int_0^\infty 4 \pi a^2 \pi  B_\lambda(T(r)) Q_\lambda d\lambda~,
\label{bilancio}
\end{equation}
where $F^S_\lambda (r)$ is the stellar flux, $Q_\lambda$ is the grain absorption efficiency,
$T(r)$ the dust temperature, $ B_\lambda(T(r))$ the black body emissivity  and $a$ the dust grain size. 
This calculation assumes that the heating by non radiative processes
and by the diffuse radiation field is negligible so that we limited ourselves to compute 
$Q_\lambda$ for a typical particle size distribution \cite{Mathis}
with optical constants derived by Draine and Lee \cite{Draine}.
Specifically, the numerical value of $R_h$ is obtained by using equation (\ref{bilancio})
with $T(R_h)=T_h$.
Assuming $T_h \simeq 1300$ K, we show  in Fig. \ref{fig1} the { ratio $R_h/\sqrt{R_S}$ as} a function of the stellar 
surface temperature $T_S$. This figure allows us to { derive  $R_h$, once} $R_S$ and $T_S$ are given.

The explicit form of the intensities $I_+(\mu)$ and $I_-(\mu)$ is given in Refs.
\cite{Simmons02,Ignace06}.
It turns out that the polarization $P$ linearly depends on 
the total optical depth $\tau = n_h \sigma R_h/(\beta-1)$, 
where $\sigma$ is the scattering cross-section and the { scatterers}
 are taken to exist only for $r>R_h$.
An estimate of the order of magnitude of $\tau$ 
is derived assuming a stationary, spherically symmetric stellar wind 
%
%
\cite{Ignace08} 
\begin{eqnarray}
\tau  = 2 \times 10^{-3} \eta {\mathcal K} 
\left(\frac{\dot{M}} { 10^{-9}~M_{\odot} {\rm yr}^{-1} }\right)
\left(\frac {30~{\rm km ~ s}^{-1}} {v_{\infty}}\right)
\left(\frac {24R_{\odot}} {R_h} \right)~,
\label{tau} 
\end{eqnarray}
where $\eta \simeq 0.01$ is the dust-to-gas mass density ratio, 
${\mathcal K} \simeq 200$ cm$^2$ g$^{-1}$ is the dust opacity at
$\lambda > 5500$ \AA,
$\dot{M}$ is the mass-loss rate and $v_{\infty}$ the asymptotic wind velocity.


To estimate $\tau$ for cool giant stars,
we relate $\dot{M}$ to the stellar parameters of the magnified star. 
Indeed, it is well known that from main sequence to AGB phases, 
$\dot{M}$ increases by 7 order of magnitude \cite{Ferguson}.
By performing numerical simulations of the mass loss of intermediate and 
low-mass stars, it was  shown that $\dot M$ obeys to the 
relation 
\begin{eqnarray}
\dot{M} = 2 \times 10^{-14}  
\frac{ (L/L_{\odot} ) ( R/R_{\odot} )^3 
( T/T_{\odot})^9 }  
{(M/M_{\odot})^2}~ M_{\odot}~{\rm yr}^{-1}~.
\label{dotM}
\end{eqnarray}
For the more common stars evolving
from main sequence to red giant star phases,
$\dot{M}$ values in the range $(10^{-13} - 10^{-8})~M_{\odot}~{\rm yr}^{-1}$ 
are expected. This corresponds, from Eq. (\ref{tau}) to values of $\tau$ in the range 
$4 \times 10^{-7} - 4 \times 10^{-2}$.

\begin{table*}
\centering
\caption{Best-fit parameters of the event OGLE-2011-BLG-0950/MOA-2011-BLG-336
for the binary (A) and  planetary (B) lens models.}
\medskip
\begin{tabular}{c|c|c|c|c|c|c|c|}
\hline
model            & $t_0$   & $u_0 (10^{-3}$) & $t_E$  & $d$   &  $q$                & $\alpha$ & $\rho_S (10^{-3})$   \\
\hline
                 & (days)  &                &  (day) &       &                     &          &                     \\
\hline
    A            & 5786.40 &  9.3           & 61.39  & 0.075 & 0.83                & 0.739    & 3.2                 \\
    B            & 5786.40 &  8.6           & 65.21  & 0.70  & $5.8 \times 10^{-4}$& 4.664    & 4.6                 \\
\hline
\end{tabular}
\label{table}
\end{table*}

\section{Results}

\begin{figure}[htbp]
 \vspace{6.0cm} \includegraphics{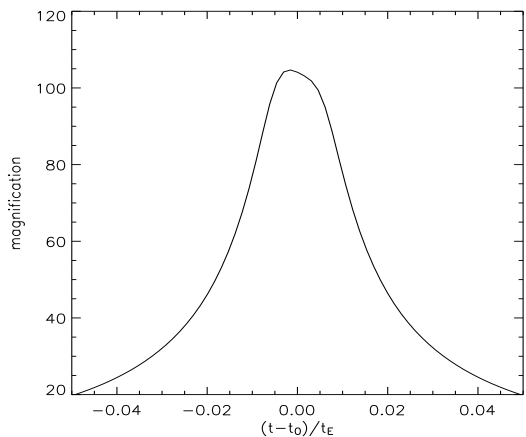}
\caption{ Simulated light-curves of the OGLE-2011-BLG-0950/MOA-2011-BLG-336 event,
corresponding to the models A and B in Table \ref{table}.
The light curves of the two models are almost identical and thus indistinguishable
and agree very well with the experimental points of the event \cite{Choi12}.}
\label{curvadiluce}
\end{figure}

In the following we focus in particular on the event OGLE-2011-BLG-0950/MOA-2011-BLG-336.
This high magnification event presents central perturbations in the light curve
that may be caused either by a binary lens (model A) or a planetary lens (model B) \cite{Choi12}.
However, by simply fitting the light curve, it is not possible to distinguish between the two solutions
and this gives rise to a specific degeneracy in the parameter space.
In Figs. 1 and 2 of the above mentioned paper the degeneracy of the solutions is fully described.
Despite the basically different caustic shapes and the resulting magnification patterns 
of the two solutions, 
the source trajectory in both cases is crossing (with different angles) the regions of {\it negative} 
perturbation in such a way that the morphology of the resulting perturbations are the same
\footnote{{\it Negative} perturbation means that the magnification of the perturbed part 
of the light curve is lower than the magnification of the corresponding single-lensing 
event. In the model A, the source trajectory passes the negative perturbation region
behind an arrowhead-shaped central caustic, in the model B the analogous region is 
between two cusps of an astroid-shaped caustic \cite{Choi12}.}.
The best-fit model parameters are summarized in Table \ref{table} and 
the simulated { event light-curves (almost identical for the two models A and B) are shown in the Fig. \ref{curvadiluce}.}

\begin{figure}[htbp]
\vspace{9cm} \includegraphics{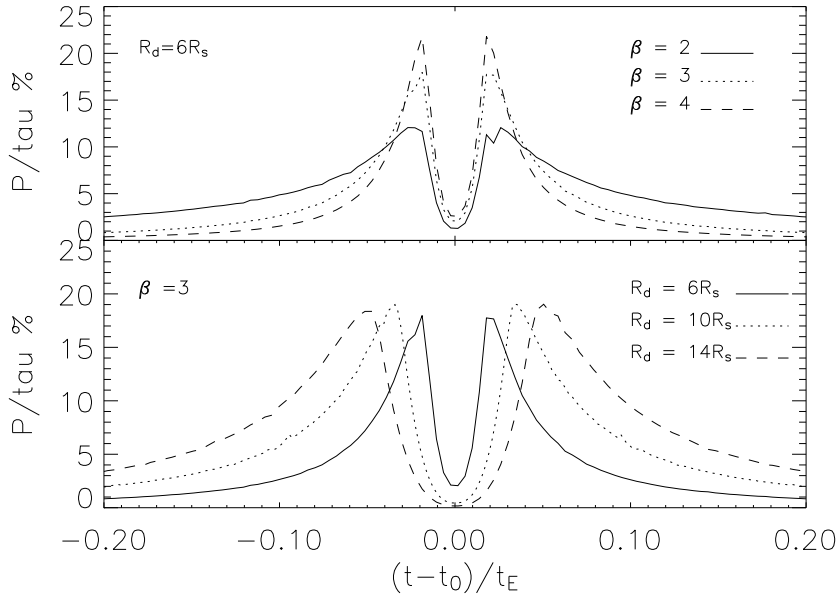}
\caption{Polarization profiles in units of $\tau$ are given for different
values of $\beta$ and $R_h$. In the upper panel, assuming $R_h=6~R_S$, the continuous, dotted and dashed
curves correspond to $\beta=2,~3,~4$, respectively. In the bottom panel with $\beta=3$, we vary 
$R_h=6~R_S$ (continuous), $R_h=10~R_S$ (dotted) and $R_h=14~R_S$ (dashed line).}
\label{fig3}
\end{figure}
In Fig. \ref{fig3}, for the model A,  we show { simulated}
 polarization curves $P(t)$ (in units of $\tau$) evaluated by fixing  
the best-fit binary parameters given in Table \ref{table} and varying the polarization model parameters $\beta$ and $R_h$ values. 
A typical polarization curve has two maxima and one minimum, bracketed by the maxima, which coincides with the instant 
$t_0$ of maximum amplification. Similar results (not shown) are obtained for the model B. The polarization signal gets the 
maximum when the condensation radius $R_h$ (the radius of the central cavity in the stellar atmosphere) enters and exits 
the lensing region. Two peaks appear at symmetrical position with respect to $t_0$ and the characteristic time scale 
$\Delta t_h$ between them is related to the {\it transit} duration of the central cavity
\begin{equation}
\Delta t_h \simeq 2 t_E \times \sqrt{R_h^2-u_0^2}~.
\label{transittime}
\end{equation}

In Fig. \ref{fig3} (upper panel, where $R_h = 6 ~R_S$), we explore the effect on the polarization signal
of varying the parameter $\beta$. As one can see, the maximum polarization value increases with increasing $\beta$. 
This behavior is expected since the dust density gradient across $R_h$ (which is {\it transiting} the lensing region) 
increases with increasing $\beta$ and this has the effect to reinforce  the asymmetry across the stellar atmosphere which, 
ultimately, is at the origin of the polarization signal for cool giant stars. 
The bottom panel of Fig. \ref{fig3}, where we fix $\beta=3$, shows that for increasing $R_h$ values
the distance between the two maxima increases. The effect is present in the polarization curves (not shown) 
evaluated for different $\beta$ values.
\begin{figure}[htbp]
\vspace{7.5cm} \includegraphics{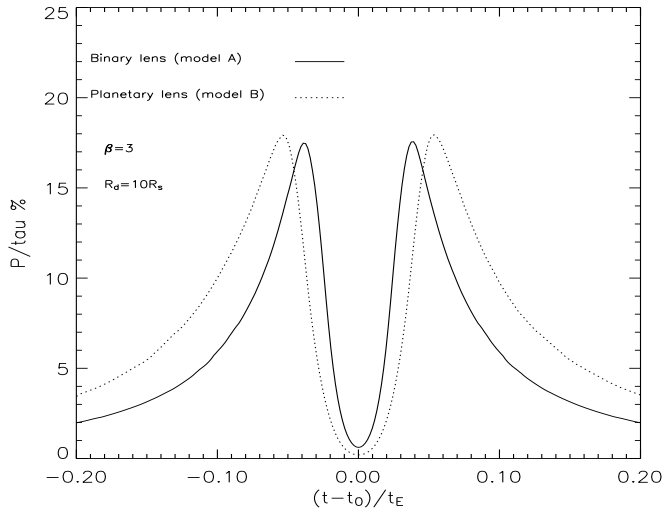}
\caption{Polarization profiles in units of $\tau$ 
for the models A and B in Table \ref{table}. 
The polarization parameters are $\beta=3$ and $R_h=10~R_S$.}
\label{fig4}
\end{figure}
{From these results it is 
evident that the only relevant model parameter is $R_h$, 
which is directly related to the observable time interval 
$\Delta t_h$, as shown in Eq. (\ref{transittime}).  
The parameter $\beta$, related to the wind acceleration mechanism, 
remains instead largely undetermined, since it does not exist an observable
uniquely related to it.}

In Fig. \ref{fig4}, by taking as an illustration $\beta=3$ and $R_h=10~R_S$, we compare the expected polarization
profiles for the best-fit models A and B in the OGLE-2011-BLG-0950/MOA-2011-BLG-336 event. 
The position of the two maxima is different and such difference remains for any selected values of 
$\beta$ and $R_h$. Therefore, an independent determination of $R_h$, based (as shown in Fig. \ref{fig1}) 
on a direct observation of the source star temperature $T_S$  and the determination of the source radius 
$R_S$ (through the best-fit to the event light-curve), allows us to distinguish the two models A and B,
namely the   binary  or planetary solution to the lens system.

{Actually, in the case of the considered event OGLE-2011-BLG-0950/MOA-2011-BLG-336,
the radius and the surface temperature of the source are unconstrained by observations 
and therefore our analysis of the simulated polarization profiles remains an exercise 
that, anyhow, shows the potentiality of the method.}

We emphasize that the detection of the polarization signals in forthcoming  microlensing events is technically 
reachable, as already noted in \cite{Ingrosso12}. However, to that aim, it is necessary to select, among all 
microlensing events, the class of the highly magnified events that also show large finite size source effects, 
in particular events with source stars belonging to the class of cool, {giant stars \cite{Zub}}. These evolved stars have 
both large radii (giving rise to relevant finite size source effects) and large stellar atmospheres, where the 
light get polarized by photon scattering on dust grains. For these events, hopefully, the dust optical dept 
$\tau$ could be $\simeq 10^{-2}$ so that the polarization signals 
$P \simeq 0.2 \%$ could be detectable, due to both the high brightness at maximum and the large time duration 
of the polarization signal. This observational programme may take advantage
of the currently available surveys plus follow up strategy already routinely used for microlensing monitoring 
towards the Galactic bulge (aimed at the detection of exoplanets). In particular, this allows one to predict 
in advance for which events and at which exact time instant the observing resources may be focused to make intensive polarization 
measurements.

We conclude by noting that polarization measurements in a binary microlensing event (OGLE-2012-BLG-0798) have been 
performed recently. The data analysis, with the aim of distinguish among the several models that give a good 
fit of the observed light curve, is at present in progress \cite{Bozza}.

\end{document}